\documentclass[a4paper, amsfonts, amssymb, amsmath, reprint, showkeys, nofootinbib, twoside]{revtex4-2}
\usepackage[english]{babel}
\usepackage[utf8]{inputenc}

\usepackage[pdftex, pdftitle={Article}, pdfauthor={Author}]{hyperref} 

\usepackage{graphicx}
\usepackage{dcolumn}
\usepackage{bm}
\usepackage{siunitx}
\usepackage{pstricks}
\usepackage{comment}
\usepackage{booktabs}
\usepackage{setspace}
\usepackage{ragged2e} 
\usepackage{upgreek}

\begin{document}

\author{Anica Hamer}
 \affiliation{%
 Physikalisches Institut, Universität Bonn, Nussallee 12, 53115 Bonn, Germany
 }
\author{Frank Vewinger}%
\affiliation{%
 Institut für Angewandte Physik, Rheinische Friedrich-Wilhelms-Universität Bonn, Bonn, Germany
 }%
\author{Michael H. Frosz}
\affiliation{%
 Max-Planck-Institut für die Physik des Lichts, Erlangen, Germany
 }%
\author{Simon Stellmer}
 \email{stellmer@uni-bonn.de}
\affiliation{%
 Physikalisches Institut, Universität Bonn, Nussallee 12, 53115 Bonn, Germany
 }%
 
\date{\today}

\DeclareSIUnit\bar{bar}
\DeclareSIUnit\cps{cps}

\title{Frequency conversion in a hydrogen-filled hollow-core fiber: power scaling, background, and bandwidth}

\begin{abstract}
Large-area quantum networks based on optical fibers allow photons at near-infrared wavelengths to travel with minimal loss. Quantum frequency conversion is a method to alter the wavelength of a single photon while maintaining its quantum state. Most commonly, nonlinear crystals are employed for this conversion process, where near-unity conversion efficiency at high fidelity has been demonstrated. Still, the crystal-based conversion process is plagued by strong background noise, very limited spectral bandwidth, and inhomogeneous temperature profiles at strong pump fields.

In previous work, we have demonstrated frequency conversion in hydrogen-filled hollow-core fibers and claimed that this conversion process does not compromise performance at strong pump fields, is essentially free of background noise, and intrinsically broadband. Here, we demonstrate that these three claims are justified.
\end{abstract}

\maketitle

\section{Introduction}
Efficient and state-preserving single photon frequency conversion is essential for the realization of quantum computing and communication in hybrid architectures, where subsystems are connected via optical links  \cite{Kimble2008,Ruetz2017,Zaske2012,Krutyanskiy2017,Fisher2021,Bell2017,Zhou2014,Ikuta2011,DeGreve2012,Tyumenev2022,Winzer2018,Deutsch2023}.
Prominent approaches to frequency conversion are based on the $\chi^{(2)}$ nonlinearity of crystals \cite{Ruetz2017, Zaske2012, Fisher2021, Bell2017, Zhou2014, Ikuta2011,DeGreve2012} and the $\chi^{(3)}$ nonlinearity of atomic gases in combination with short-pulse lasers \cite{Eramo1994,Tamaki1998,Babushkin2008,Cassataro2017,Tyumenev2022}.

Despite the recent developments around the $\chi^{(2)}$-nonlinearity of crystals that led to a near-unit conversion efficiency and a drastically reduced incoherent background, the $\chi^{(3)}$-nonlinearity of molecular gases provides significant advantages as this approach sidesteps the limitations of the very narrow acceptance bandwidth, polarization-dependence of the crystals, absorption, possible heat management, and unintended background. The potential of a $\chi^{(3)}$-nonlinearity in gases for coherent Stokes and anti-Stokes Raman scattering (CSRS/CARS) processes has been highlighted e.g. in Refs.~\cite{Boyd2008,Shipp2017,Rigneault2018,Groessle2020,Tian2023}.

In this Letter, we present broadband frequency conversion from the wavelength range of \SIrange{859}{863}{\nano\meter} to the telecom O-band at \SIrange{1336}{1350}{\nano\meter} based on the CSRS process inside a commercially available antiresonant-reflecting hollow-core fiber (ARR-HCF). Compared to previous work on a slightly different fiber \cite{Hamer2024a}, we increased the pump power by two orders of magnitude to observe a four-orders-of-magnitude improvement in efficiency. We present broadband conversion across a range of 10\,nm, and we comment on the heat management and the generation of background. Different to many other works, we employ cw pump light to retain the spectral purity of the probe photon.

The ARR-HCFs, a sub-group of the hollow-core photonic crystal fiber (HC-PCF), are based on the inhibited coupling as its guiding mechanism \cite{Belardi2019}. Well-known implementations are the revolver-type fibers or its further development, the nodeless anti-resonant fiber (NAF). Such fibers are used in this work. They are characterized by a small number of capillaries arranged around the hollow core at equal azimuthal distances within a jacket capillary, without contact of the inner capillary \cite{Ando2019, Murphy2023, Novoa2024}. For an overview of current research using these fibers, especially for four-wave-mixing processes \cite{Russell2014, Bahari2022,Afsharnia2024, Tyumenev2022,Mridha2019,Cui2023,Couny2007}, as well as a presentation of the advantages and developments of HCF, we refer to a recent review \cite{Novoa2024}.


\section{Experimental setup}
We perform frequency conversion in a band centered around \SI{863}{\nano\meter} to a band centered around \SI{1346}{\nano\meter}, a frequency range which is of interest as it allows the conversion of two entangled photons from the biexciton emission of indium arsenide (InAs)/gallium arsenide (GaAs) quantum dots \cite{Gao2012,Liu2019,Wang2019,Joecker2019} to the telecom O-band while preserving their entanglement.

\begin{figure}[tpb]
	\centering
	\includegraphics[width=\linewidth]{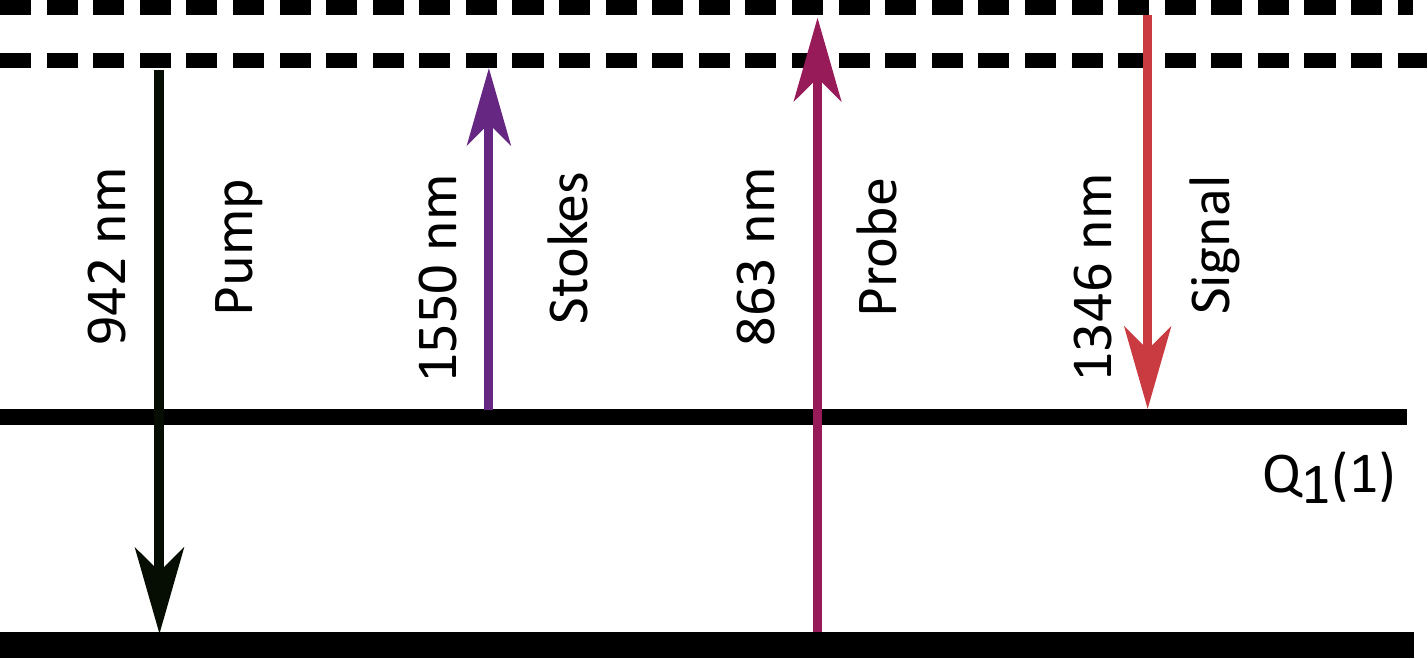}	
	\caption{The schematic diagram of the CSRS process.}
	\label{fig:CSRS}
\end{figure}

To this end, the Stokes field at \SI{1550}{\nano\meter} is adjusted to match the pump field at \SI{942}{\nano\meter} for a beating at the vibrational transition at \SI{125}{\tera\hertz} within the $Q_1(1)$ branch of molecular hydrogen. In the following, both lasers will be called pump fields. The induced nonlinear polarization is used for the conversion of the probe field at \SI{863}{\nano\meter} to the signal light at \SI{1346}{\nano\meter}. A schematic diagram of the conversion process is shown in Fig.~\ref{fig:CSRS}. 

The experimental setup has been described elsewhere \cite{Hamer2024,Hamer2024a} and is slightly modified. In Ref.~\cite{Hamer2024a}, the fiber was mounted between two high pressure chambers, and we observed a loss in pressure along the fiber. Here, the fiber is glued into a fiber holder which is placed inside a home-built, stainless steel, hermetically sealed, cylindrical high-pressure chamber using an O-ring seal. The vessel is tested up to \SI{700}{\bar}. AR-coated sapphire windows are sealed to the vessel using O-rings  \cite{Aghababaei2023, Hamer2024}.

For increased pump power at \SI{938}{\nano\meter}, we employ a vertical-external-cavity surface-emitting laser (VECSEL) at \SI{942}{\nano\meter} with a power up to \SI{7}{\watt}. To match the molecular transition, Stokes light at \SI{1550}{\nano\meter} is generated by a diode laser combined with a fiber amplifier. 

\begin{figure}[htpb]
	\centering
	\includegraphics[width=\linewidth]{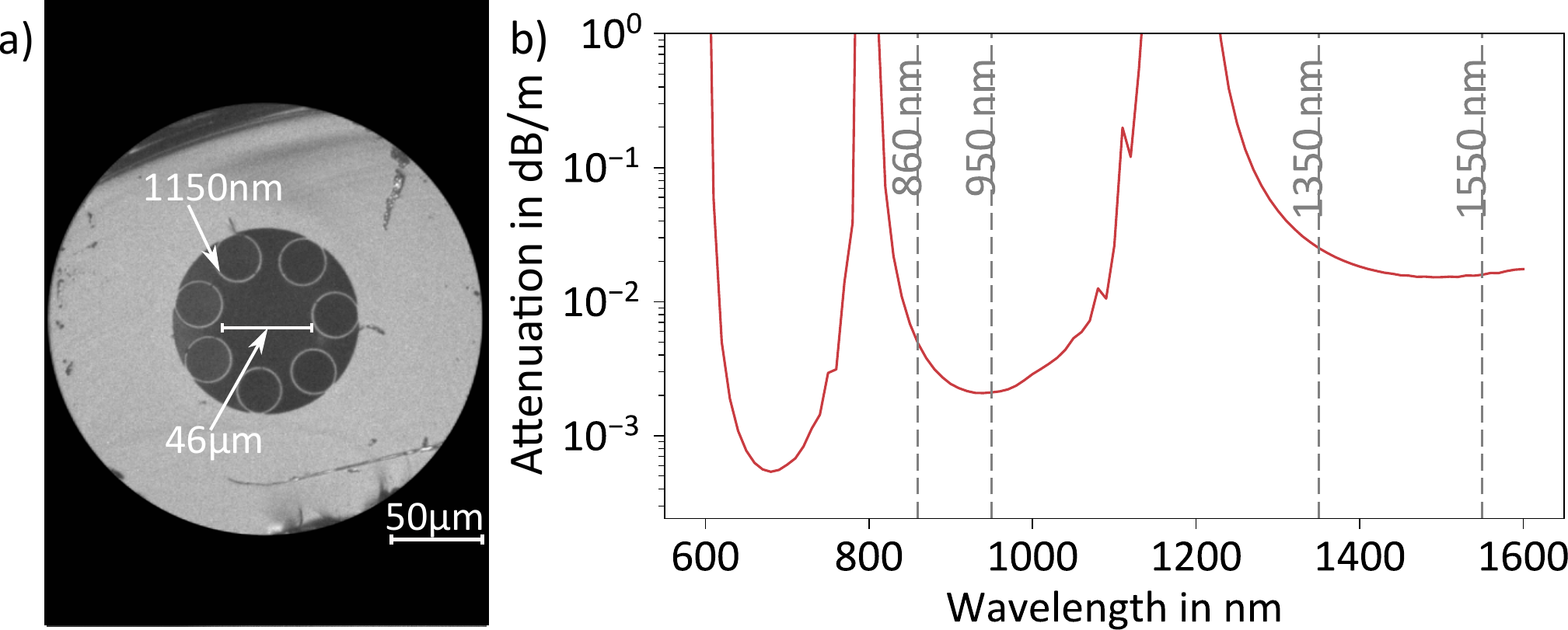}	
	\caption{a) Bright-field microscope image of the fiber facet. b) \textit{Comsol} loss calculation (red) for the wavelengths used in this work (dashed gray lines).}
	\label{fig:modes}
\end{figure}

Furthermore, we here employ a commercially available NAF ARR-HCF fiber designed for \SI{1550}{\nano\meter}. The core has a diameter of \SI{46}{\micro\meter} and is surrounded by seven capillaries with a wall thickness of \SI{1150}{\nano\meter}, as shown in Fig.~\ref{fig:modes} a). The fiber is \SI{6}{\centi\meter} long. To calculate fiber properties such as the mode field diameter or the transmission bandwidth depending on the chromatic dispersion, we use the model given by Zeisberger-Hartung-Schmidt \cite{Zeisberger2018} to obtain the effective refractive index $n_\text{eff}$ inside the fiber core with small computational effort. To support our calculations, we additionally use the finite element simulation software \textit{Comsol}, with which we can calculate the mode profiles of the Gaussian-like HE$_{11}$ fiber modes (analogous to Ref.~\cite{Hamer2024a}) and the leakage loss, see Fig.~\ref{fig:modes} b).   

These calculations are important to confirm that the fiber can support the relevant wavelengths, even though it is designed only for \SI{1550}{\nano\meter}. The coupling between the core modes and the capillaries shows pronounced resonances and incurs strong losses \cite{Tyumenev2022,Zeisberger2018}. The position of the resonances are determined by the wall thickness $w$ of the capillaries. Here, the $w$ is chosen such that the loss is minimal at the wavelengths used in the experiment. Other loss types such as imperfection losses, material absorption of the fiber \cite{Yu2016}, absorption by the hydrogen molecules \cite{Bahari2022} or bend losses are negligible here. We achieve incoupling efficiencies of \SIrange[]{72}{84}{\percent} for our wavelengths in the Gaussian-like HE$_{11}$ fiber mode for powers exceeding \SI{5}{\watt}.


\section{Frequency Conversion}\label{sec:freq}

Based on recent results \cite{Hamer2024a}, we identified the need to address pressure losses, to enhance incoupling efficiency, and to increase the power of the pump fields. We place the fiber inside a high-pressure vessel to counteract gas leakage of the fiber seals and to ensure uniform pressure over the length of the fiber. 

The photon conversion rate is pressure-dependent \cite{Rigneault2018} and scales as
\begin{equation}\label{eq:I}
    I_\text{1346} \propto {| \chi^{(3)}(\omega) |}^2 L^2 {\mathrm{sinc}^2}\left(\frac{\Delta \beta (p) \cdot L }{2}\right) I_\text{942} I_\text{1550} I_\text{863},
\end{equation}
where $\chi^{(3)}(\omega)$ denotes the third-order nonlinear susceptibility and $L$ the interaction length. 
Here, $\Delta \beta = -\beta_\text{942nm} +\beta_\text{1550nm}+\beta_\text{863nm}-\beta_\text{1346nm}$ is the phase matching condition based on the propagation constant $\beta = 2 \pi \nu/c \cdot  n_{\text{eff}}(p,\lambda)$ which itself includes the pressure $p$ and the wavelength $\lambda$ dependent effective refractive index $n_{\text{eff}}$ from the Zeisberger model \cite{Zeisberger2018}.

As a first step to optimize the conversion efficiency, we scan the pressure of the hydrogen gas and find an optimal pressure near \SI{60}{\bar}. This optimum is independent of pump powers. All following measurements are conducted at a pressure of \SI{60}{\bar}.  


\subsection{Increase of pump power} \label{sub:intensity}

\begin{figure}[hptb]
	\centering
	\includegraphics[width=\linewidth]{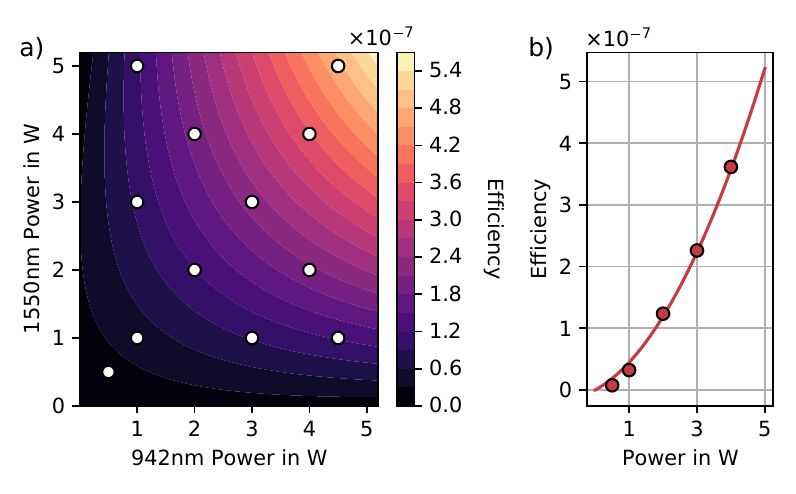}
         \caption{a) Conversion efficiency in dependence of the pump powers, color-coded in a surface plot with a 2D second-order polynomial fit applied to the data points (white symbols). b) A cut through the plane at $I_\text{942} = I_\text{1550}$.} 
	\label{fig:surface}
\end{figure}

In this experiment, we increase the power of the two pump fields by a factor of 100 compared to previous work \cite{Hamer2024a}. We measure the efficiency of the process across various combinations of pump powers, the results are presented as a surface plot in Fig.\ref{fig:surface} a). As expected, we observe maximum conversion efficiency at the highest pump powers. The conversion efficiency is color-coded and plotted as a function of the power in the two pump fields. A 2D second-order polynomial fit (analogous to Eq.\ref{eq:I}) is applied to the data (white symbols), yielding a maximum internal efficiency of \SI{5.2e-7}{}. This efficiency accounts for detector and optical losses. In all measurements, the probe power is set to \SI{1}{\micro\watt}.

For the case of equal pump powers, $I_\text{942} = I_\text{1550}$, the data and the fitted conversion efficiency are plotted in Fig.~\ref{fig:surface} b), showing the quadratic behavior and the potential of the experiment. The conversion efficiency here is limited by the maximal power of the lasers. The coupling efficiency into the fiber is very stable and independent of pump power, indicating a good thermal contact between the fiber and its mount, as well as an even distribution of heat along the fiber.


\subsection{Background noise}

In crystals, frequency conversion often includes undesired Raman scattering, which can emit into the spectral range of the converted photons. These scattering processes are intrinsic to the crystal material. However, these effects are significantly suppressed when using a gas-filled hollow-core fiber due to the minimal overlap between the glass structure and the light fields \cite{Tyumenev2022}.

In the following, we quantify the background noise in our conversion process. The detectors are operated at \SI{10}{\percent} detection efficiency. Without any light fields on, we observe a dark count rate of \SI{270}{\cps}, which includes detector dark counts and ambient stray light.

With only the \SI{1550}{\nano\meter} laser turned on, we observe an increased count rate that scales linearly with the laser power. The count rate strongly depends on the position of certain optical elements, and can be reduced through shields and baffles.
The count rate is independent of the gas pressure inside the fiber. We conclude that these counts, which amount to about \SI{1500}{\cps} at \SI{5}{\watt} pump power, are caused by stray light from the laser and optics. Dedicated shielding will minimize this background further.

A non-trivial contribution to the background noise originates from the \SI{942}{\nano\meter} laser, where Raman processes can up-shift the wavelength of the photons to pass the 1350-nm filter in front of the detectors. Indeed, at full power of \SI{5}{\watt}, we observe an increase by about \SI{2500}{\cps}. These photons are not polarized. The count rate increases linearly for power levels up to \SI{3.5}{\watt}, beyond which we observe a threshold-type behavior and a steep increase in count rate. Also, we observe that the count rate depends strongly on the mode matching of the \SI{942}{\nano\meter}-light into the fiber. For optimized coupling and below-threshold pump power, count rates are on the order of a few \SI{100}{\cps}, but increase by orders of magnitude if light is coupled into the capillaries or cladding structure. These two observations suggest Raman scattering with its typical threshold behavior \cite{Jahns2001} occurring within the fiber, likely due to increased interaction with the glass material. Such a Raman process may involve the conversion of light from \SI{942}{\nano\meter} to approximately \SI{1346}{\nano\meter}, attributed to the presence of hydroxyl groups \cite{Brooke2016, Noll2020} in the silicon-based glass \cite{Putz2012} analogous to conventional optical fibers. 

Further improvements in fiber coupling will reduce this background further. In summary, no background noise was detected that would correlate to the pressure of the hydrogen gas or to the \SI{863}{\nano\meter} light.


\subsection{Bandwidth}
\begin{figure}[hptb]
	\centering
	\includegraphics[width=\linewidth]{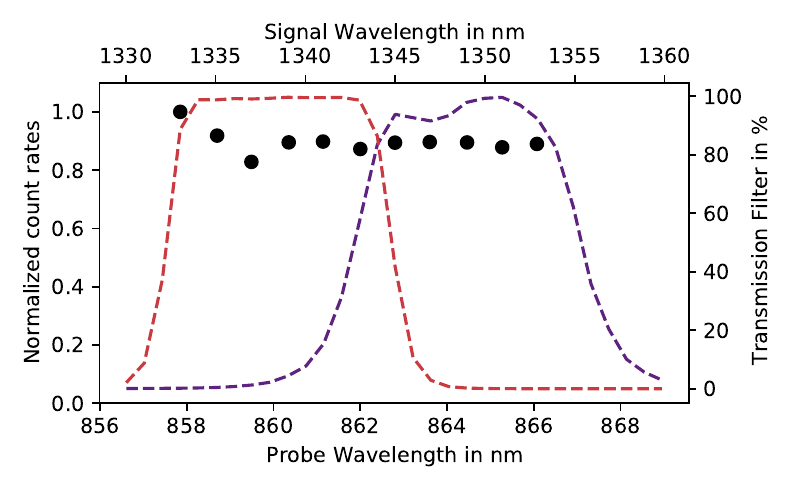}
         \caption{Normalized count rates of the converted probe light (black data points) in dependence of the probe wavelength around \SI{863}{\nano\meter} converted to around \SI{1346}{\nano\meter}. Count rates are corrected for the transmission values of two different filters (dashed red and purple lines). Statistical error bars are smaller than the symbol size.
         }
	\label{fig:broad}
\end{figure}

We test the bandwidth of the CSRS conversion, having in mind the typical distribution of emission wavelengths from individual quantum dots or the 3-nm wavelength difference between the two entangled photons from the bi-exciton decay in InAs/GaAs quantum dots \cite{Hamer2022}. Here, we step the wavelength of the probe light from \SIrange[]{858}{866}{\nano\meter} and record the conversion rate. To reduce the background, we employ two different bandpass filters around \SI{1340}{\nano\meter} and \SI{1350}{\nano\meter}, each with a width of \SI{12}{\nano\meter}. Conversion rates are corrected for filter transmission and shown in Fig.~\ref{fig:broad}. Indeed, the conversion efficiency is constant (standard deviation: \SI{2.6}{\percent}) across a wavelength range of about \SI{10}{\nano\meter}. Our measurement range is limited by the availability of suitable filters, and we expect the bandwidth to be much larger. This bandwidth needs to be compared to crystal-based frequency conversion, where the bandwidth is on the order of \SI{0.1}{\nano\meter}. Our approach demonstrates high suitability for simultaneous conversion of photons at different wavelengths, or very short photons with a large spectral spread.

\section{Conclusion}
To conclude, we have demonstrated frequency conversion in high-pressure hydrogen with a hollow-core fiber with an internal conversion efficiency of \SI{5.2e-5}{\percent}, corresponding to an increase by 4.5 orders of magnitude that was achieved through stronger pump fields. We studied potential sources of background noise, where no contributions related to the hydrogen gas were found. 

To further enhance the conversion efficiency, we plan to implement a longer fiber, leveraging the quadratic dependence of the conversion rate on the interaction length (Eq.~\ref{eq:I}). The coherence required to reach the maximal conversion efficiency \cite{Tyumenev2022} could not build up in the short length of the fiber used here. The fiber will be enclosed in a pressure-stable pipe to ensure uniform pressure along the length of the fiber.

{Funding - }
We acknowledge funding by Deutsche Forschungsgemeinschaft DFG through grant INST 217/978-1 FUGG and through the Cluster of Excellence ML4Q (EXC 2004/1 – 390534769), as well as funding by BMBF through the QuantERA project QuantumGuide.

{Acknowledgments - }
We thank all members of the Cluster of Excellence ML4Q and the QuantumGuide collaboration, especially Thorsten Peters, as well as Philipp Hänisch, for stimulating discussions. 

{Disclosures - }
The authors declare no conflicts of interest.

{Data Availability Statement - }
Data underlying the results presented in this paper are not publicly available at this time but may be obtained from the authors upon reasonable request.


\bibliography{bib}

\end{document}